\documentclass[fleqn,usenatbib]{mnras}

\usepackage{newtxtext,newtxmath}
\usepackage[T1]{fontenc}
\usepackage{ae,aecompl}

\usepackage{amsmath}
\usepackage{amssymb}
\usepackage{epic}
\usepackage{epstopdf}
\usepackage{graphicx}
\usepackage{hyperref}
\usepackage{multirow}
\usepackage{natbib}
\usepackage{overpic}
\usepackage{color}
\AtBeginShipout{%
  \ifnum\value{page}>1 %
    \typeout{* Additional boxing of page `\thepage'}%
    \setbox\AtBeginShipoutBox=\hbox{\copy\AtBeginShipoutBox}%
  \fi
}

\title[Duty-cycle and energetics of remnant radio-loud AGN]{Duty-cycle and energetics of remnant radio-loud AGN}

\author[R. J. Turner]{
Ross J. Turner$^{1,2}$\thanks{Email: turner.rj@icloud.com}\\
$^{1}$School of Physical Sciences, University of Tasmania, Private Bag 37, Hobart, 7001, Australia\\
$^{2}$CSIRO Astronomy and Space Science, Post Office Box 76, Epping, New South Wales 1710, Australia}

\date{Accepted 2018 February 14. Received 2018 February 13; in original form 2017 November 29}

\pubyear{2018}

\begin{document}

\label{firstpage}
\pagerange{\pageref{firstpage}--\pageref{lastpage}}
\maketitle

\begin{abstract}

Deriving the energetics of remnant and restarted active galactic nuclei (AGNs) is much more challenging than for active sources due to the complexity in accurately determining the time since the nucleus switched-off. I resolve this problem using a new approach that combines spectral ageing and dynamical models to tightly constrain the energetics and duty-cycles of dying sources. 
Fitting the shape of the integrated radio spectrum yields the fraction of the source age the nucleus is active; this, in addition to the flux density, source size, axis ratio, and properties of the host environment, provides a constraint on dynamical models describing the remnant radio source. 
This technique is used to derive the intrinsic properties of the well-studied remnant radio source B2\,0924+30. This object is found to spend $50_{-12}^{+14}\rm\, Myr$ in the active phase and a further $28_{-5}^{+6}\rm\, Myr$ in the quiescent phase, have a jet kinetic power of $3.6_{-1.7}^{+3.0}\times10^{37}\rm\, W$, and a lobe magnetic field strength below equipartition at the $8\sigma$ level. The integrated spectra of restarted and intermittent radio sources is found to yield a `steep-shallow' shape when the previous outburst occurred within $100\rm\, Myr$. The duty-cycle of B2\,0924+30 is hence constrained to be $\delta < 0.15$ by fitting the shortest time to the previous comparable outburst that does not appreciably modify the remnant spectrum. The time-averaged feedback energy imparted by AGNs into their host galaxy environments can in this manner be quantified. 

\end{abstract}


\section{INTRODUCTION}
\label{sec:INTRODUCTION}

The supermassive black holes (SMBHs) at the centres of active galaxies and the properties of their surrounding interstellar and intergalactic medium are known to be closely linked \citep{McNamara+2012}. However, quantifying the energetics of this active galactic nuclei (AGN) feedback requires knowledge of the energy imparted in each outburst and the duty-cycle of the SMBH activity.
The relativistic jets fuelling the expansion of {a pair of synchrotron-emitting plasma lobes} terminate once the accretion of matter onto the SMBH ceases. 
{The bright radio-frequency core and jets quickly disappear upon the cessation of the fuel supply, though the lobes remain visible for some time in these remnant sources \citep{Slee+2001}.}
The integrated {radio spectrum arising from the lobes} steepens sharply beyond some break frequency due to the preferential radiating of high-energy electrons \citep{Kardashev+1962, Pacholczyk+1970, Jaffe+1973, Komissarov+1994}. The shape of the spectrum is further modified if the jet restarts, injecting a fresh batch of electrons {and inflating another pair of radio lobes} \citep{Murgia+2011}. 
The age of remnant radio galaxies, and the duty-cycle in the case of restarted sources, is thus normally derived based on the shape of their radio spectra \citep[and an assumed lobe magnetic field strength; e.g.][]{Jamrozy+2004, Murgia+2011, Konar+2013b, Brienza+2016, Shulevski+2017}. The evolution of the synchrotron spectra of these dying radio sources has been explored extensively in the literature \citep[e.g.][]{Komissarov+1994, Kaiser+2002}.

Remnants can {in many cases be} discovered in radio surveys by examining sources exhibiting a steep spectral index\footnote{{The spectral index, $\alpha$, is defined in this work by $S = \nu^{-\alpha}$ for flux density $S$ and frequency $\nu$.}} (i.e. $\alpha > \alpha_{\rm inj} + 0.5$, where the injection spectral index is typically in the range $0.5 < \alpha_{\rm inj} < 1$ {and related to the energy spectrum of electrons injected into the lobe, $N(E)dE \propto E^{-s}dE$, through $\alpha_{\rm inj} = (s - 1)/2$}). {However, the steep spectral index selection criterion (typically $\alpha > 1.3$) is expected to produce a biased sample of remnants since this technique excludes: remnants observed only at frequencies below the spectral turnover (where the spectrum is flatter); recently switched off sources \citep[e.g. 3C28,][]{Harwood+2017}; and remnants with non-standard spectral shapes \citep[see blob1,][]{Brienza+2016}.}
{Steep-spectrum radio sources comprise at most ten to twenty percent of lobed FR-Is in the Lockman Hole field \citep{Brienza+2017}, observed using Low-Frequency Array (LOFAR) and Westerbork Synthesis Radio Telescope (WSRT). Mock catalogues confirm that classifying remnants based on the spectral steepness will detect only a fraction of the population unless high frequency observations ($>$$1.4\rm\, GHz$) are included.}  {\citet{Hardcastle+2016} further provide an upper limit to the remnant fraction of $30\%$ by looking for current core activity in high-resolution Very Large Array (VLA) images of radio sources observed using LOFAR in the Herschel-ATLAS North Galactic Pole field. {The remnant fraction is tightened to $<$$9\%$ by \citet{Mahatma+2018} who used higher sensitivity and spatial resolution follow-up observations to detect compact cores in some of the candidate remnants.}}

The spectral signatures of intermittent (i.e. repeatedly dying and restarting) and singly restarted sources are less well-defined, however these objects present an opportunity to directly measure the duty-cycle. These sources may be identified by measuring the spectral index across a low- and high-frequency band \citep[e.g.][]{Callingham+2017}.
The double-double radio galaxies examined by \citet{Konar+2013b} and \citet{Orru+2015}, for example, are found to spend minimal time in the inactive phase between outbursts. {The currently active radio jets thus overrun the plasma from the previous outburst on short timescales and merge with the outer remnant lobes \citep{Konar+2013a}.}
By contrast, remnants show ages of order $100\rm\, Myr$ and are typically observed soon after the jet switches-off \citep{Jamrozy+2004, Murgia+2011, Shulevski+2017}, with few remnants in the quiescent phase longer than the active phase \citep[but see][]{Brienza+2016}. Measurements of the population duty-cycle are thus impeded by both the ability to detect the oldest remnants and to recognise the spectra of restarted sources.

The Radio AGN in Semi-analytic Environments \citep[RAiSE;][]{Turner+2015, Turner+2017a} model for the dynamical and synchrotron evolution of active radio galaxies is extended in this work to consider remnant and restarted sources. The existing form of the model has been shown to: (1) yield jet kinetic powers consistent with X-ray inverse-Compton measurements; (2) reproduce the surface brightness and spectral age distributions of 3C31 and 3C436 respectively; and (3) have dynamical evolution in the lobe length, volume and axis ratio consistent with hydrodynamical simulations \citep{Turner+2017b}.
The extended RAiSE model for remnant and restarted sources strengthens the previous investigations of dying radio sources by combining both the spectral and dynamical constraints on the intrinsic properties of these objects.

In this paper, I model the radio spectra of remnants and restarted sources to aid the identification of these objects from their integrated emission in large-sky surveys (Section \ref{sec:RADIO SPECTRAL FITTING}). The effect the duty-cycle has on the shape of the spectrum and the ability to detect previous outbursts is investigated in Section \ref{sec:RESTARTING RADIO SOURCES}. The RAiSE radio source evolution model is extended in Section \ref{sec:DYNAMICAL AND EMISSIVITY MODELS} enabling the dynamics of remnant and restarted sources to be modelled, and is applied to the well-studied remnant B2\,0924+30 in Section \ref{sec:REMNANT PARAMETER FITTING} to test the extension to the model. The duty-cycle and kinetic power of this remnant radio galaxy are derived to investigate the energetics of AGN feedback.

The $\Lambda \rm CDM$ concordance cosmology with $\Omega_{\rm M} = 0.3$, $\Omega_\Lambda = 0.7$ and $H_0 = 70 \rm\,km \,s^{-1} \,Mpc^{-1}$ \citep{Komatsu+2011} {is} assumed throughout the paper.

\section{REMNANT SPECTRAL AGEING MODEL}
\label{sec:RADIO SPECTRAL FITTING}

The spectral age of radio AGNs can be derived from the steepening in the observed synchrotron emission spectrum due to the synchrotron and inverse-Compton loss processes \citep[e.g.][]{Jaffe+1973, Myers+1985, Alexander+1987, Murgia+1999, Jamrozy+2008}. 
\citet{Turner+2017b} showed that the continuous injection model \citep[CI model;][]{Kardashev+1962, Pacholczyk+1970}, which assumes a constant injection of fresh particles over the active lifetime of the source, well fits the radio spectra of {over $86\%$} of FR-IIs in the \citet{Mullin+2008} sample of 3C sources {over a broad frequency range (typically 12 measurements between 0.01 and $10\rm\, GHz$ from \citealt{Laing+1980})}. This work only considers the simple, albeit most common, case of a radio source observed during its active phase; \citet{Komissarov+1994} presented an extension to the CI model for remnant radio sources, the CI-off (also known as the KGJP) model. {\citet{Harwood+2017} similarly examine the spectra of five lobed radio sources (including one remnant) 
but find only two are well fitted by the CI (or CI-off) model; their spectra include high precision broad bandwidth measurements and observations from several authors, so despite the statistical rejection, the general shape of the remaining three sources can still be well characterised by the CI model. However, the continuous injection model assumption of a time-invariant magnetic field strength (in particular) is violated in the lobes of radio AGNs since the field strength decreases as the source expands adiabatically \citep[e.g.][]{Turner+2015}. Regardless, \citet{Turner+2017b} find that the CI model provides a good empirical fit to simulated spectra \citep[which include magnetic field evolution; see][]{Turner+2017a} of lobed radio sources (but not `flaring jet' FR-Is, e.g. 3C31), and accurately retrieves the source dynamical age.}

The standard CI model is a function of frequency $\nu$, the energy of the synchrotron-emitting electrons $E$, their pitch angle $\xi$, the magnetic field strength $B$, and the spectral age of the source $\tau$. However, \citet{Turner+2017b} find that the shape of the CI model is only weakly affected by the assumed magnetic field distribution, and that its strength merely translates the spectrum in the $\log\nu$--$\log S$ plane. The magnetic field strength and spectral age are degenerate parameters for translations along the frequency axis, {and the flux density estimated by spectral ageing models does not} implicitly include adiabatic losses or consider the evolutionary history of the magnetic field strength. \citet{Turner+2017b} instead recast the standard CI model in terms of a dimensionless function of the frequency, field strength and energy:

\begin{equation}
x = \frac{\nu}{\nu_{\rm b}} = \frac{4 \pi {m_{\rm e}}^3 c^4 \nu}{3 e E^2 B \sin \xi} ,
\end{equation}

where $\nu_{\rm b}$ is the spectral break frequency beyond which (i.e. at higher frequencies) the spectrum steepens due to the loss mechanisms, $e$ is the electron charge, $m_{\rm e}$ is the electron mass, and $c$ is the speed of light.

Following the method of \citet{Turner+2017b}, the standard integral representation of the CI and CI-off model {flux densities} is reduced from a triple to a double integral in terms of only $x$ and the pitch angle, with the field strength as a constant scaling factor. That is,

\begin{equation}
J(\nu) = J_0 \nu^{-s/2} \int_0^{\pi/2} \sin^{(s + 4)/2} \xi \int_0^\infty F(x) \mathcal{N}(x) dx d\xi ,
\end{equation}

where $J_0$ is a frequency-independent constant, $s$ is the exponent of the electron energy distribution at injection, $F(x)$ is the single-electron synchrotron radiation spectrum, and $\mathcal{N}(x)$ is the frequency-dependent component of the electron energy distribution. This component of the electron energy distribution is modified from that of the standard CI model \citep[Equation 10 of][]{Turner+2017b} using the CI-off model expressions derived by \citet{Komissarov+1994}:

\begin{equation}
\mathcal{N}(x) = x^{-1/2}
\begin{cases}
(x^{1/2} - \varsigma^{1/2})^{s - 1} - (x^{1/2} - \iota^{1/2})^{s - 1} &\iota < x \\
(x^{1/2} - \varsigma^{1/2})^{s - 1} &\varsigma \leqslant x \leqslant \iota \\
0 &x < \varsigma
\end{cases} ,
\end{equation}

\begin{figure}
\begin{center}
\includegraphics[width=0.48\textwidth]{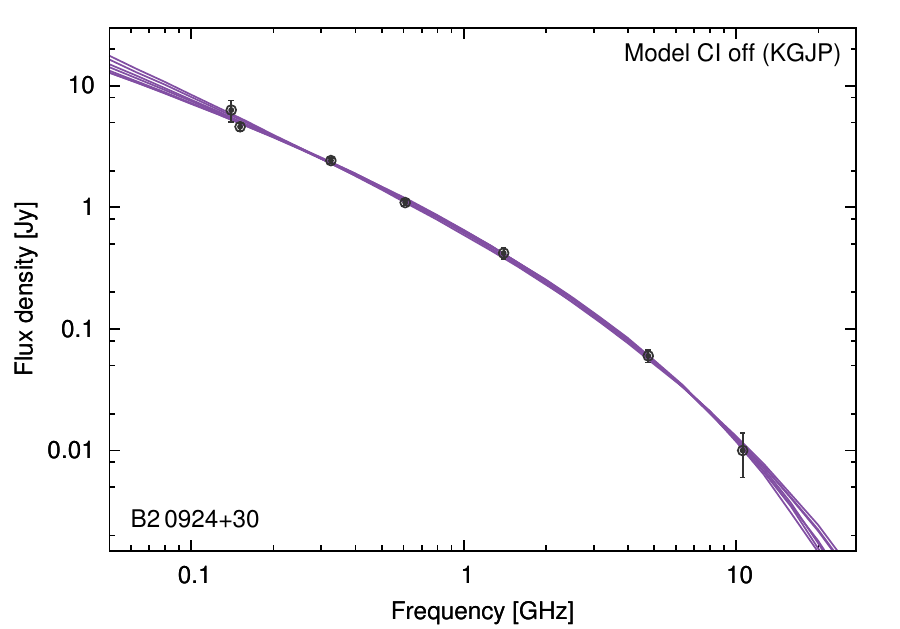} 
\end{center}
\caption{Fits of the CI-off model to the observed spectrum of B2\,0924+30 assuming various electron injection indices between $s = 2.4$ and 3. The flux density measurements are taken from \citet{Shulevski+2017} and plotted with $1\sigma$ errorbars. The limited frequency coverage precludes the injection index, break frequency and active fraction from being simultaneously constrained.}
\label{fig:spectralfits}
\end{figure}

where $\iota(\nu, \xi) = \nu/(\nu_{\rm b} \sin \xi)$ and $\varsigma(\nu, \xi) = \iota(\nu, \xi) (t_{\rm off}/\tau)^2$ for time spent in the quiescent phase $t_{\rm off} = \tau - t_{\rm on}$; we define $T = t_{\rm off}/\tau$ as the the quiescent fraction in this work. The CI-off model converges to the standard form of the CI model described in Equation 10 of \citet{Turner+2017b} if the radio source spends no time in the quiescent phase (i.e. the spectral age equals the active age of the source, $\tau = t_{\rm on}$). Fits of the CI-off model to the spectrum of B2\,0924+30 (see discussion in Section \ref{sec:REMNANT PARAMETER FITTING}) are shown in Figure \ref{fig:spectralfits} for a range of electron injection indices, $s$.
Finally, the active age, quiescent fraction, break frequency and lobe magnetic field strength can be related through

\begin{equation}
\tau = \frac{t_{\rm on}}{1 - T} = \frac{\upsilon B^{1/2}}{B^2 + {B_{\rm ic}}^2} \left[\nu_{\rm b} (1 + z) \right]^{-1/2} ,
\label{spectralage}
\end{equation}

where $B_{\rm ic} = 0.318(1+z)^2\rm\, nT$ is the magnetic field equivalent to the cosmic microwave background radiation, $z$ is the redshift of the remnant, and the constant of proportionality $\upsilon$ is defined in Equation 5 of \citet{Turner+2017b}. These parameters can be uniquely constrained by also considering the lobe dynamics using the RAiSE radio source evolution model.

\section{RESTARTED AND INTERMITTENT SOURCES}
\label{sec:RESTARTING RADIO SOURCES}

The spectra of restarted radio sources are a logical extension to the remnant radio galaxy models presented in the previous section. These sources can be thought of as a remnant {radio galaxy into which} a fresh outburst {has been injected} from the active nucleus; i.e. their spectra is described by the sum of a CI-off and standard CI model component. This approach is somewhat simplistic as it does not consider any evolution in the magnetic field strength of the remnant between outbursts or instabilities which may collapse the lobe; these effects are considered in the full RAiSE model (Section \ref{sec:DYNAMICAL AND EMISSIVITY MODELS}) and are expected to only be especially problematic for the longest quiescent timescales.

\subsection{Energetics of restarted radio sources}
\label{sec:Duty-cycle and energetics of restarted sources}

{The duration of each AGN outburst is dependent on the availability of gas to fuel the black hole, whilst the jet kinetic power is a function of the black hole mass, spin and accretion rate \citep[e.g.][]{Meier+2001}.} 
{The lossless integrated flux density arising from a given outburst is related to the active age and jet power, $Q$, through a modified form of Equation 12 of \citet{Alexander+2000}:}

\begin{equation}
J(\nu) = J_1\, Q^\frac{22 + 2s - (s + 5)\beta}{4(5-\beta)} {t_{\rm on}}^{\!\frac{16 - 4s - (s + 5)\beta}{4(5-\beta)}}
\label{fluxpower}
\end{equation}

{where $s = 2\alpha_{\rm inj} + 1$ is again the injection index of the electron energy distribution, $\beta$ is the exponent of the power law external density profile (i.e. $\rho \propto r^{-\beta}$), and $J_1$ is a frequency-dependent constant of proportionality. For typical values of $s = 2.4$ and $\beta = 1$, the flux density relationship simplifies to $J \propto Q^{1.2} {t_{\rm on}}^{\!-0.35}$. The flux density of outbursts with durations differing by a factor of a few are therefore expected to be somewhat similar in the lossless case. The radiative loss processes reduce the emission in older (active) sources, however this will also occur in the remnant phase; the spectra from previous outbursts are therefore expected to look quite similar at the present time irrespective of their active age. This is further confirmed numerically by summing the contributions from several outbursts of differing duration simulated using the CI-off and standard CI spectral ageing models.}

{The integrated flux density is quite sensitive to variations in the jet power of each outburst (see Equation \ref{fluxpower}); more powerful outbursts yield radio spectra normalised to higher flux densities. The observed spectra of restarted radio sources are therefore expected to show no signature of earlier outbursts if the present outburst has a much higher power than the previous AGN activity. The spectra of radio sources with an earlier, more powerful outburst is that of a remnant with some flattening at high-frequencies from the more recent, less powerful outbursts; the spectra of such sources are left to be investigated on a case-by-case basis.}
{However, observations of double-double radio sources suggest the jet power in successive outbursts from a given AGN are often very similar. \citet{Marecki+2016} found the jet powers of the inner and outer doubles of J1706+4340 are within 10\%.} {Further, \citet{Konar+2013a} argue that similar injection spectral indices observed in the inner and outer doubles of $75\%$ of the radio sources in their sample are due to similar jet powers in the two outbursts (based on the tight correlation between jet power and injection index).}
{However, similar jet powers can not be expected for successive outbursts in every source; in particular, each outburst triggered by cold gas accretion during galaxy mergers and interactions will have very different characteristics.} 


\subsection{Spectra of restarted radio sources}

\begin{figure}
\begin{center}
\includegraphics[width=0.48\textwidth]{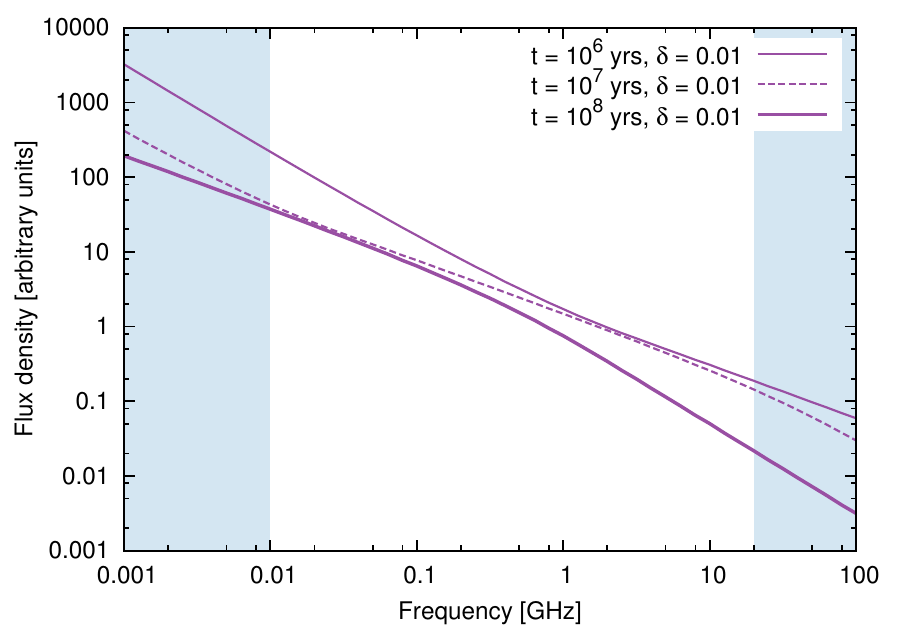} 
\end{center}
\caption{Simulated radio spectra for multiple AGN outbursts each producing radio lobes with field strength $0.5\rm\, nT$ at $z = 0.1$. Three curves are shown assuming active lifetimes of 1, 10 and $100\rm\, Myr$ each for a duty-cycle of $\delta = 0.01$ and injection index of $s = 2.4$. The shading indicates frequencies at which radio observations are not typically available.}
\label{fig:restarting}
\end{figure}

The integrated radio spectra of a wide range of restarted sources are simulated using the CI-off and standard CI spectral ageing models. {Based on the previous discussion,} each radio source is assumed to maintain an identical active lifetime, $t_{\rm on}$, duty-cycle, $\delta = t_{\rm on}/(t_{\rm on} + t_{\rm off})$, and {lobe} magnetic field strength {(function of active age, jet power and environment\footnote{{The environment is disturbed by each AGN outburst, only returning towards its hydrostatic equilibrium on the dynamical timescale of the cluster (much longer than the duty-cycle of restarted sources). However, the identical duty-cycle and energetics assumed for each outburst will lead to comparable environments between each episode of SMBH activity.}})}, $B$, between outbursts for the dynamical timescale of the cluster ($\sim 1\rm\,Gyr$). The kinetic-mode of AGN activity remains the dominant mechanism well beyond this timescale \citep{Fabian+2012}, {and all but {those} sources with the shortest active lifetimes and the highest duty-cycles will have at most a few outbursts contributing to their observed radio spectrum}.
The modelled spectra for three sources with different active lifetimes are shown in Figure \ref{fig:restarting} for a typical duty-cycle of $\delta = 0.01$ \citep{Best+2005}. The oldest radio source ($t_{\rm on} = 100\rm\, Myr$) has a standard broken power-law spectrum with $\alpha_{\rm inj} = 0.7$ at low frequencies steepening to $\alpha = 1.2$ after the break. However, the youngest modelled source ($t_{\rm on} = 1\rm\, Myr$), with a correspondingly shorter quiescent phase for the same duty-cycle, has a steep spectrum at low frequencies before flattening to $\alpha_{\rm inj} = 0.7$; synchrotron-emitting electrons from the previous outburst $100\rm\, Myr$ prior contribute preferentially to the emissivity at low frequencies. Restarted radio sources can therefore produce a `steep-shallow' spectrum in contrast to the typical `shallow-steep' synchrotron spectrum. This high-frequency flattening of an aged spectra by a fresh outburst is seen in the observations of WNB1829+6911 and B2\,0120+33 examined by \citet{Murgia+2011}.

\subsection{Exploring spectral index parameter space}

The shape of restarted radio source spectra is further investigated by measuring their spectral index at both low and high frequencies. The low frequency spectral index, $\alpha_{\rm low}$, is defined between the standard observing frequencies of $151\rm\, MHz$ (e.g. LOFAR and MWA) and $1.4\rm\, GHz$ (e.g. ASKAP) with the high frequency index, $\alpha_{\rm high}$, between $5.5$ and $9.5\rm\, GHz$ (e.g. ATCA). The radio spectra for a wide range of restarted sources are simulated by convolving the spectral outputs of the CI and CI-off ageing models. Radio galaxies are simulated with active ages of between 1 and $100\rm\, Myr$ \citep{Turner+2015, Turner+2017b}, duty-cycles in the range $\delta = 0.001$ to 0.3 \citep{Best+2005}, and with magnetic field strengths ranging from 0.1 up to $10\rm\, nT$ \citep{Ineson+2017}. Some of these restarted radio sources are likely too faint to be readily detected by current telescopes but are included to present a complete interpretation of the parameter space.

\begin{figure}
\begin{center}
\includegraphics[width=0.48\textwidth]{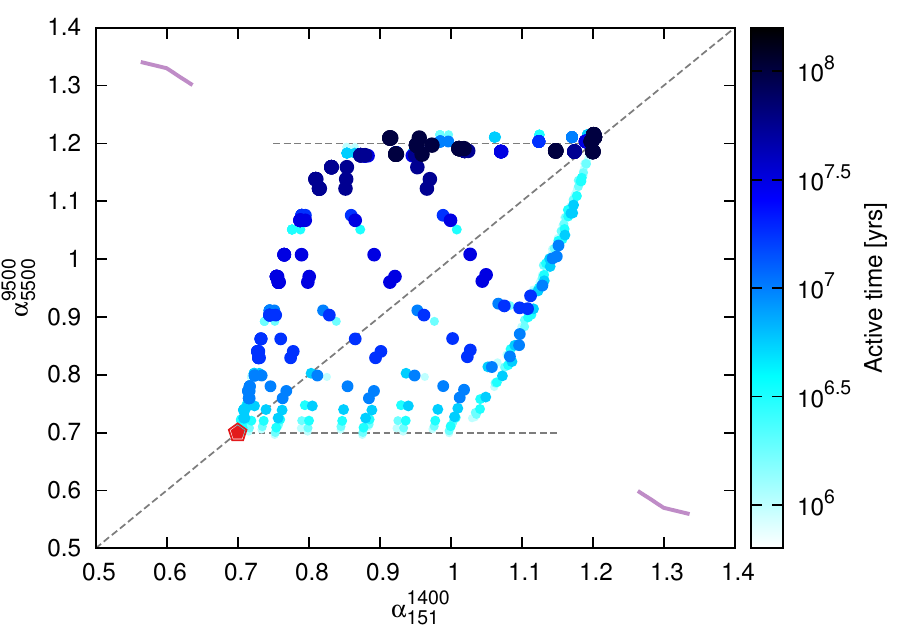}\\ \includegraphics[width=0.48\textwidth]{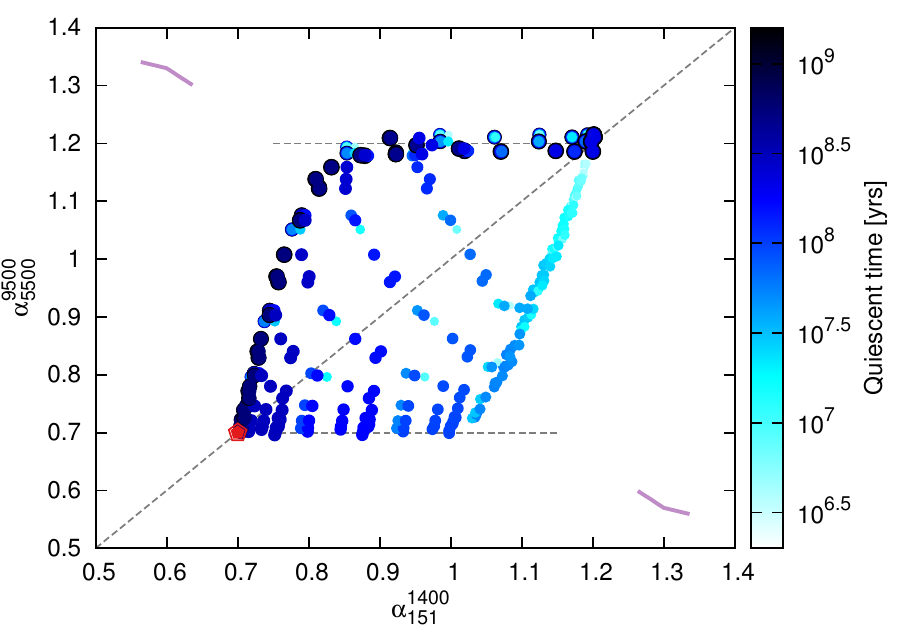}
\end{center}
\caption{Spectral indices fitted to the spectra of modelled remnant and restarted sources for a range of magnetic field strengths, and active and quiescent lifetimes. The spectral indices are measured across both a low frequency ($151\rm\, MHz$ to $1.4\rm\, GHz$) and a high frequency ($5.5$ to $9.5\rm\, GHz$) band to characterise the spectral shape. The top panel shows the spectral indices of the simulated radio sources in $\alpha_{\rm low}$--$\alpha_{\rm high}$ parameter space colour coded by the active time of each outburst (the point size is also scaled by the active time). The injection spectral index of $\alpha_{\rm inj} = 0.7$ is marked by a red pentagon, {and extended along the horizontal axis with a dashed line; the steepest aged spectrum ($\alpha = \alpha_{\rm inj} + 0.5$) is similarly shown with a dashed line. Finally, the solid purple lines show an indicative spectral shape for sources on each side of the one-to-one line (dashed diagonal line).} The bottom panel similarly shows the distribution colour coded by the quiescent timescale between each outburst.}
\label{fig:alphaalpha}
\end{figure}

The spectral indices fitted to the spectra of these modelled restarted sources are shown in Figure \ref{fig:alphaalpha} in $\alpha_{\rm low}$--$\alpha_{\rm high}$ parameter space. The active lifetime of the source, $t_{\rm on}$ and the time between outbursts, $t_{\rm off}$, are shown in the figure through both the point size and colour. The vast majority of simulated sources lie on one of two `hysteresis-like' curves from $(\alpha_{\rm low},\alpha_{\rm high}) = (0.7,0.7)$ to $(1.2,1.2)$; sources lying on the upper track have a `shallow-steep' spectrum and those on the lower curve a `steep-shallow' spectrum. 
\citet{Callingham+2017} find a similar clustering in observations of radio galaxies taken as part of of the GaLactic and Extragalactic All-sky Murchison Widefield Array (GLEAM) survey; their Figure 2.
The radio sources with the typical `shallow-steep' synchrotron spectrum have the longest quiescent timescales, generally in excess of $1\rm\, Gyr$\footnote{Shorter quiescent timescales may yield the same result if the lobe from the previous outburst has collapsed due to instabilities, hence removing any contribution from the remnant to the emissivity.}. The spectral index steepens as these sources age until it reaches a maximum value of $\alpha_{\rm inj} + 0.5 = 1.2$; the radiative loss mechanisms are more effective at high frequencies leading to the spectra steepening first at high frequencies then at low frequencies as the source ages. By contrast, the objects exhibiting a `steep-shallow' spectrum generally have a previous active phase within the last $100\rm\, Myr$, and consequently have shorter active lifetimes. These restarted radio sources are generally too short-lived for their spectra to steepen over the course of a single outburst, however earlier outbursts may show significant spectral ageing. 

The radio sources occupying the lower `steep-shallow' curve in Figure \ref{fig:alphaalpha} can be separated into two further populations: (1) singly restarted radio sources, and (2) `intermittent' radio sources where the jet fuelling switches on and off periodically. The spectrum of a singly restarted radio source is the convolution of a power-law (young restarted component) and a low-frequency contribution from the much older remnant \citep[see observations of][]{Murgia+2011}; these objects lie on the $\alpha_{\rm high} = 0.7$ line in the figure. By contrast, intermittent radio sources may have several injections of synchrotron-emitting electrons contributing to the radio spectrum. Each outburst is short (e.g. accretion of a gas cloud) but regular in effect leading to a time-averaged jet power, $\bar{Q}$, describing the dynamics of a single lobe inflation across several outbursts. The spectra of these presently-active intermittent radio sources differs considerably from the standard CI model since oldest synchrotron-emitting electrons from each outburst cycle (i.e. during each quiescent phase) are not present. By contrast, intermittent radio sources seen in an inactive phase will have their spectra quickly approach one more closely resembling that of a remnant. Radio sources exhibiting a typical active-phase morphology but which are poorly fitted by a standard CI spectrum may well be explained through fuelling by intermittent outbursts.

\section{DYNAMICAL AND EMISSIVITY MODELS}
\label{sec:DYNAMICAL AND EMISSIVITY MODELS}

The spectral modelling of remnant and restarted radio sources in the previous section provides a good qualitative description of their expected signature in multifrequency radio observations. The accurate modelling of individual objects requires a more sophisticated treatment considering the adiabatic loss process in the integrated luminosity calculation, the evolution of the lobe magnetic field strength, and the Rayleigh-Taylor mixing of the remnant with the host environment. Remnant radio lobes may also rise buoyantly along the density profile ultimately forming `mushroom cloud' shaped plumes. 

The \citet{Turner+2015} dynamical and \citet{Turner+2017a} synchrotron emissivity models for lobed FR-II/I radio sources include all radiative loss mechanisms, magnetic field strength evolution, and the end-of-life lobe mixing. These models are easily modified to include the coasting phase upon the jet switching off, or alternatively the quiescent phase between outbursts in restarted sources. The differential equations in the RAiSE radio source evolution model describing the acceleration of the lobe surface \citep[equations 7 and 8 of][]{Turner+2015} are changed to include a conditional function for the jet power:

\begin{equation}
Q(t) = Q_0
\begin{cases}
1 &t \in \mathbb{T}_{\rm on} \\
0 &t \in \mathbb{T}_{\rm off}
\end{cases} ,
\end{equation}

where $Q_0$ is the jet power in the active state, $\mathbb{T}_{\rm on}$ is the set of times at which the jet is active, and $\mathbb{T}_{\rm off}$ is the set of quiescent times.
The differential equations are valid for multiple outbursts, however if Rayleigh-Taylor instabilities have collapsed the lobe, the restarted jet would instead need to be modelled expanding into a perturbed environment. Such radio sources are beyond the scope of the analytic modelling in this work but can be examined using hydrodynamical simulations. 

The integrated luminosity calculation is similarly updated by modifying the loss function in Equation 4 of \citet{Turner+2017a} to only consider synchrotron-emitting electrons accelerated whilst the SMBH is active. That is,

\begin{equation}
\mathcal{Y}(t) = \int_{0}^{t} \frac{Q(t_{\rm i})}{Q_0} f(t, t_{\rm i}) dt_{\rm i} ,
\end{equation}

where $f(t, t_{\rm i})$ is the integrand in Equation 4 of \citet{Turner+2017a}, and $Q(t_{\rm i})/Q_0$ is the instantaneous jet power scaled by its active value (i.e. 0 or 1 in this work).

\begin{figure}
\begin{center}
\includegraphics[width=0.45\textwidth]{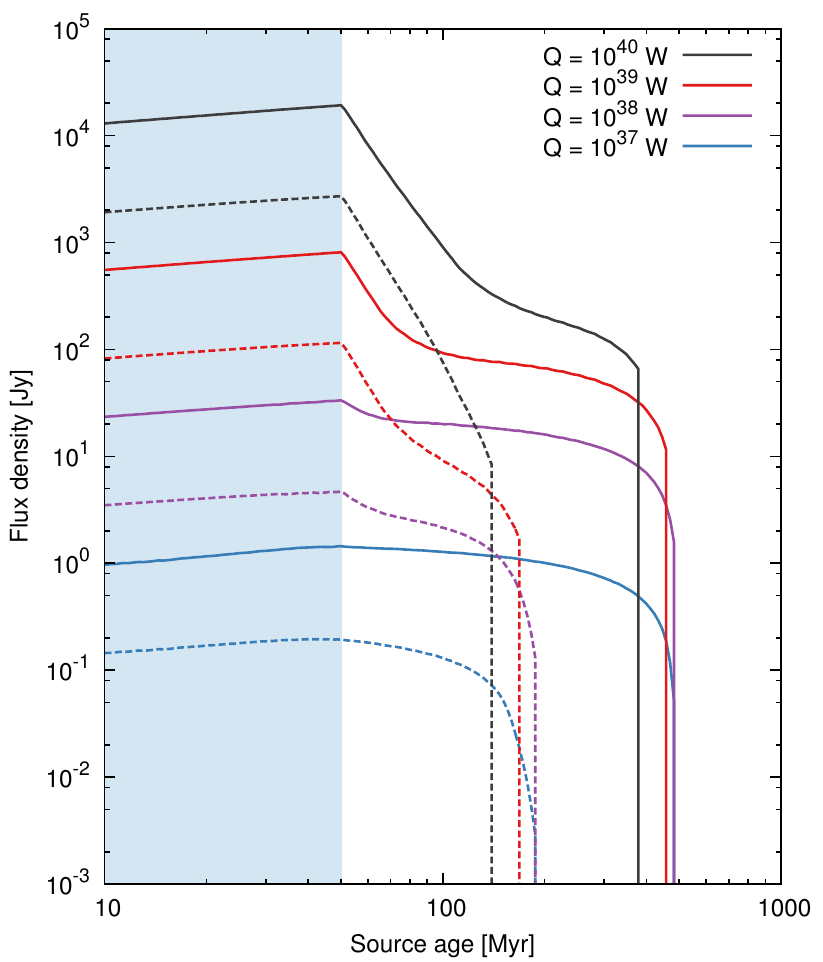} 
\end{center}
\caption{Flux density of a `B2\,0924+30-like' radio source for four jet powers as a function of the source age. These tracks are modelled assuming a host gas density of $\rho_{\rm x} = 7\times10^{-26}\rm\, kg\,m^{-3}$, a injection energy index of $s=2.7$ \citep{Shulevski+2017}, an equipartition factor of $B/B_{\rm eq} = 0.38$ \citep[population average, see][]{Turner+2017b}, and an active lifetime of $50\rm\, Myr$ (see Section \ref{sec:REMNANT PARAMETER FITTING}). The flux density tracks are plotted for two standard observing frequencies; $151\rm\, MHz$ (solid) and $1.4\rm\, GHz$ (dashed). The blue shading indicates the duration of the active phase.}
\label{fig:remnantpower}
\end{figure}

Simulated age--luminosity tracks for a `B2\,0924+30-like' remnant with a single $50\rm\, Myr$ outburst (see Section \ref{sec:REMNANT PARAMETER FITTING}) are shown in Figure \ref{fig:remnantpower} for a range of jet powers. The luminosity at both the $151\rm\, MHz$ and $1.4\rm\, GHz$ standard observing frequencies investigated here falls off rapidly after the jet is switched-off for the high-powered sources. The majority of the energy in the electron population is stored in the most freshly injected electrons; the luminosity quickly reduces once the jet no longer replenishes this population. By contrast, the lowest-powered source maintains its luminosity long after entering the quiescent phase. This source comes into pressure equilibrium with the host environment preventing any appreciable adiabatic expansion of the electron packets injected after that time. Most of the energy in low-powered sources may therefore be stored in the older synchrotron-emitting electrons. 
Finally, the sharp drop in the $1.4\rm\, GHz$ luminosity at approximately $200\rm\, Myr$, and the $151\rm\, MHz$ luminosity at $500\rm\, Myr$, occurs because there are no longer any electrons in the injected population capable of emitting at the observing frequency. The radio lobe is likely to have been deformed by buoyant forces well before this time, so observations of this effect are not to be expected.

The synchrotron emissivity model is further used to test the applicability of the CI-off model for remnant sources. The spectra of the simulated sources in Figure \ref{fig:remnantpower} are found to be well fitted by the CI-off model at all times after the jet switches off over the $100\rm\, MHz$ to $10\rm\, GHz$ frequency range. Further, the spectra simulated for restarted and intermittent sources are consistent with the qualitative findings in Section \ref{sec:RESTARTING RADIO SOURCES}.

\section{INTRINSIC PARAMETER FITTING}
\label{sec:REMNANT PARAMETER FITTING}

The radio spectra of remnant and restarted radio galaxies are a complicated function of the active and quiescent timescales, and the evolutionary history of the lobe magnetic field strength. Following the method of \citet{Turner+2017b}, I constrain the intrinsic properties of these radio sources based on their spectra, size and host environment. The shape of the spectra of remnants can be parametrised in terms of the injection index, $\alpha_{\rm inj}$, break frequency, $\nu_{\rm b}$, and the fraction of time spent in the quiescent and active phases, $T = t_{\rm off}/\tau$ by fitting the CI-off spectral ageing model. The spectra of restarted sources require additional parameters to fit secondary spectral components, or in the case of presently-active intermittent sources the duty cycle $\delta$ is introduced instead of the quiescent fraction.
Intrinsic parameter fitting algorithms are applied to the well-studied remnant B2\,0924+30 to test the veracity of the models developed in this work.

\subsection{Remnant radio galaxy B2\,0924+30}
\label{sec:Remnant radio galaxy B2 0924+30}

\citet{Shulevski+2017} observed the steep spectrum radio source B2\,0924+30 using the \emph{Low Frequency Array} (LOFAR) telescope, and combined with archival data, found the integrated spectrum is well fitted by the CI-off model (between $140\rm\, MHz$ and $10.55\rm\, GHz$). The radio source has a projected linear size of $360\rm\, kpc$, present-time axis ratio of $\mathcal{A} \sim 2.55$ and a morphology resembling that of a remnant \citeauthor{FR+1974} type-II \citep[FR-II;][]{Cordey+1987}. The remnant is hosted by the elliptical galaxy IC 2476 (UGC 5043) which is a member of the poor galaxy cluster WBL\,224 located at $z = 0.026141$ \citep{White+1999}. This galaxy is the brightest member of the larger Zwicky cluster 0926.5+3026 and is located at 0.39 the radius of the Zwicky cluster where the cluster density profile is expected to be locally flat. \citet{Shulevski+2017} and \citet{Jamrozy+2004} measure the equipartition magnetic field strength of the remnant as between 0.135 and $0.16\rm\, nT$ following the technique of \citet{Miley+1980}. These measurements are equivalent to a lobe pressure of $p = 7\times10^{-14}\rm\, Pa$, or an external density of $\rho_{\rm x} = 7\times10^{-26}\rm\, kg\,m^{-3}$ assuming the remnant is in pressure equilibrium with the host environment (as is the case for the fitted properties of B2\,0924+30).

The radio spectrum of remnant B2\,0924+30 is fitted using the CI-off model (Section \ref{sec:RADIO SPECTRAL FITTING}), with the best fits shown in Figure \ref{fig:spectralfits} assuming different injection indices. The limited frequency coverage, which omits observations of the power-law fall-off expected at low-frequencies, prevents each of the free parameters from being uniquely constrained with confidence. 
\citet{Shulevski+2017} instead obtain a robust measurement of the injection index by comparing the spectral shape of the emission from 41 regions across the remnant to fit the global electron energy distribution; this technique counters the effect of spatial variations in the magnetic field, electron density and radiative losses.
The different age electron populations are all consistent with an injection index of $s = 2\alpha_{\rm inj} + 1 = 2.7$, and hence this value will also be assumed here in the spectral fitting. I thus fit a break frequency of $\nu_{\rm b} = 2.3_{-0.6}^{+0.8}\rm\, GHz$ and a quiescent to active age fraction of $T = 0.36\pm0.08$. The radio source properties and spectral fits are summarised in Table \ref{tab:summary}.

\begin{table}
\begin{center}
\caption[]{Observed and derived parameters for the remnant radio galaxy B2\,0924+30. The first four rows tabulate the directly observed quantities, the next three are parameters derived from the radio spectrum, and the remaining five rows list those calculated from the observables using the RAiSE radio source evolution model. {The table includes $1\sigma$ measurement uncertainties.}}
\label{tab:summary}
\renewcommand{\arraystretch}{1.1}
\setlength{\tabcolsep}{8pt}
\begin{tabular}{ccrl}
\hline\hline 
Parameter&Symbol&\multicolumn{2}{c}{Measurement}
\\
\hline 
axis ratio (present)	&	$\mathcal{A}$&	$\sim$$2.55$\!\!\!\!\!&\\
linear size	&	$D$&	$360$\!\!\!\!\!&\!\!\!\!\!$\rm kpc$\\
luminosity ($151\rm\, MHz$)	&$L_{151}$&$\!\!\!\!\!(6.9{\small\pm0.5})\times10^{24}$\!\!\!\!\!&\!\!\!\!\!$\rm W\,Hz^{-1}$\\
external density	&	$\rho_{\rm x}$&	$(7{\small\pm1})\times10^{-26}$\!\!\!\!\!&\!\!\!\!\!$\rm kg\,m^{-3}$
\\
\hline 
break frequency	&	$\nu_{\rm b}$&	$2.3_{-0.6}^{+0.8}$\!\!\!\!\!&\!\!\!\!\!$\rm GHz$\\
injection index	&	\!$s = 2\alpha_{\rm inj} + 1$\!&	$2.7$\!\!\!\!\!&\\
quiescent fraction	&	$T = t_{\rm off}/\tau$&	$0.36{\small\pm0.08}$\!\!\!\!\!&
\\
\hline 
axis ratio (initial)	&	$A$&	$\sim$$2.1$\!\!\!\!\!&\\
equipartition factor	&	$B/B_{\rm eq}$&	$0.31{\small\pm0.05}$\!\!\!\!\!&\\
jet power (two-sided)	&	$Q$&	$3.6_{-1.7}^{+3.0}\times10^{37}$\!\!\!\!\!&\!\!\!\!\!$\rm W$\\
active lifetime	&	$t_{\rm on}$&	$50_{-9}^{+12}$\!\!\!\!\!&\!\!\!\!\!$\rm Myr$\\
source age	&	$\tau = t_{\rm on} + t_{\rm off}$&	$78_{-12}^{+14}$\!\!\!\!\!&\!\!\!\!\!$\rm Myr$
\\
\hline
\end{tabular}
\end{center}
\end{table}

The integrated spectrum of B2\,0924+30 has also been studied extensively in the literature, with several authors fitting it to estimate the time spent in the active and quiescent phases.
\citet{Jamrozy+2004} apply the Jaffe-Parola \citep[JP;][]{Jaffe+1973} model to the spectrum to fit an injection spectral index of $\alpha_{\rm inj} = 0.87\pm0.09$. They find an overall average source age of $\tau = 54\pm12\rm\, Myr$ by fitting the break frequency and estimating the equipartition magnetic field strength following the technique of \citet{Miley+1980}. \citet{Shulevski+2017} instead fit the spectrum with the CI-off model, but again assume a magnetic field strength based on equipartition arguments. They obtain active and quiescent ages of $t_{\rm on} = 55.65\pm2.25\rm\, Myr$ and $t_{\rm off} = 32.04\pm1.57\rm\, Myr$ respectively, leading to a total source age of $\tau \sim 88\rm\, Myr$.

\subsection{Bayesian parameter estimation}
\label{sec:Parameter fitting}

The most likely jet power $Q$, active age $t_{\rm on}$, and equipartition factor $B/B_{\rm eq}$ (i.e. fraction of equipartition magnetic field strength) are estimated for B2\,0924+30 using a Bayesian approach. These intrinsic parameters are constrained based on the $151\rm\, MHz$ integrated luminosity, spectral break frequency, quiescent fraction, source size, present-time axis ratio and the host galaxy environment (Table \ref{tab:summary}). The radio source evolution is simulated for a range of jet powers, active ages and equipartition factors, and taking the other observed values for the relevant model parameters. The likelihood of each simulation matching the observations is calculated following the method of \citet{Turner+2017b}, with the active age/jet power/equipartition factor triplet associated with the maximum likelihood taken as our best estimate. The uncertainty in the fitting process is reduced to less than the measurement uncertainties by simulating observations over a very fine grid with a spacing of less than $0.05\rm\, dex$ for each of the fitted parameters. 

The two-sided jet power of B2\,0924+30 is estimated to be $3.6_{-1.7}^{+3.0}\times10^{37}\rm\,W$, and the equipartition factor as $0.31\pm0.05$ corresponding to a lobe magnetic field strength of approximately $0.042\rm\, nT$ (below equipartition at the $8\sigma$ level). The active age of the remnant is fitted as $50_{-9}^{+12}\rm\, Myr$ consistent with the measurement of \citet{Shulevski+2017}. This corresponds to an energy of $9_{-4}^{+8}\times10^{52}\rm\, J$ injected by the kinetic-mode of the AGN into its surrounding environment over the duration of this outburst. The source age and quiescent timescale are directly related to the active age through the quiescent fraction, leading to estimates of $78_{-12}^{+14}\rm\, Myr$ and $28_{-5}^{+6}\rm\, Myr$ respectively. This source age estimate is consistent with those of both \citet{Jamrozy+2004} and \citet{Shulevski+2017}. 

The stability of fitting algorithms omitting one or more observables is tested but are found to be highly unstable if the spectral fits are not included. By contrast, the active and quiescent ages can be reproduced for B2\,0924+30 if only the source size is excluded; however this conclusion may not be true for general sources.

\subsection{Duty-cycle and time-averaged feedback energy}
\label{sec:Duty-cycle}

\begin{figure}
\begin{center}
\includegraphics[width=0.48\textwidth]{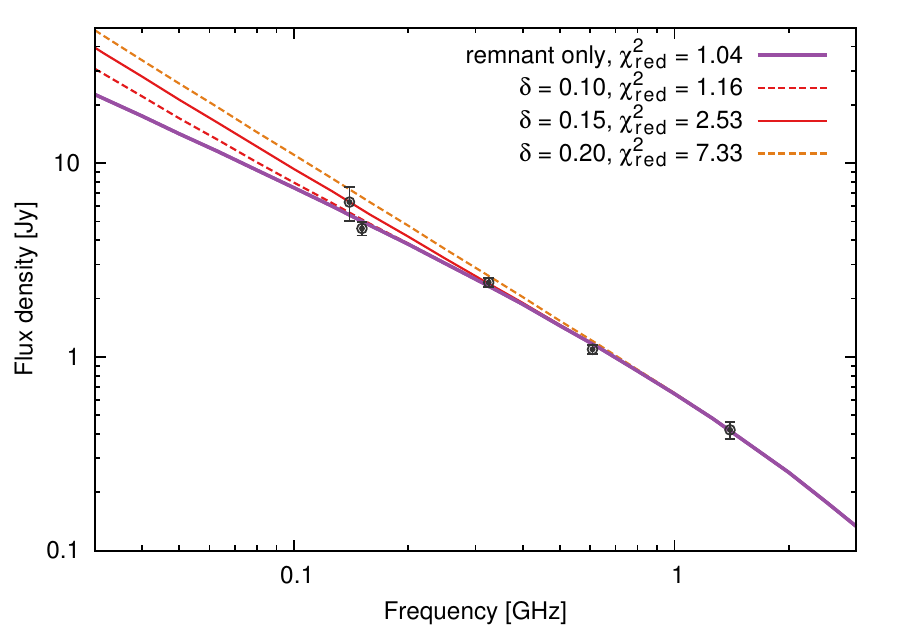} 
\end{center}
\caption{Spectral constraint on the duty-cycle of the host galaxy of remnant B2\,0924+30 (i.e. IC 2476). The remnant spectrum of B2\,0924+30 is supplemented by a spectral component from an identical outburst some time in the past, with the time between outbursts defined by the duty-cycle $\delta$. The best fit CI-off model to the remnant spectrum is plotted in solid purple; spectral models including a component from previous outbursts are shown for $\delta = 0.1$ (dashed red), $\delta = 0.15$ (solid red) and $\delta = 0.2$ (dashed orange). The reduced chi-squared statistic for each model fit to the measured spectra (black points and errorbars) is included in the legend; {the model has four degrees of freedom when fitting the full spectrum.}}
\label{fig:dutycycle}
\end{figure}

The duty-cycle of the host galaxy of remnant B2\,0924+30 (i.e. the elliptical galaxy IC 2476) can be constrained from the source age and active lifetime to be less than $\delta \lesssim 0.64$. This estimate can be improved upon by fitting the shortest time to the previous outburst that does not appreciably modify the remnant spectrum. Any previous AGN activity is assumed to have identical jet power, equipartition factor and active time to the most recent (observed) outburst {(see discussion in Section \ref{sec:Duty-cycle and energetics of restarted sources})}.
Less energetic outbursts may occur in between these two powerful outbursts (i.e. the observed and previous outburst with comparable energy) without modifying the shape of the radio spectrum, however such outbursts would also impart minimal feedback energy into AGN host environment (in comparison) and thus need not be considered. {By contrast, a more powerful outburst would modify the shape of the spectrum more appreciably than for identical jet powers, simply leading to a rather conservative upper bound on the duty cycle.}  
The possibility that the entire spectrum of B2\,0924+30 (not just the lowest frequencies) has been steepened by a previous outburst is excluded since if the measured spectral index of $\alpha = 0.85$ resulted from an aged spectral component it would necessitate an unphysical injection index of $\alpha_{\rm inj} \lesssim 0.35$ \citep[{a lower limit of $\alpha_{\rm inj} = 0.5$ is set by the assumption of first-order Fermi acceleration;}][]{Kardashev+1962, Pacholczyk+1970}.

Modified spectral ageing models including a component arising from a previous outburst are shown in Figure \ref{fig:dutycycle} and compared to the observed spectrum of B2\,0924+30. The shortest timescale to any previous comparable outburst is found to be at least $360\rm\, Myr$ (to when activity ceased), corresponding to a time between outbursts of at least $330\rm\, Myr$ at the $2\sigma$ level. The duty-cycle of the host galaxy of remnant B2\,0924+30 is therefore less than $\delta < 0.15$.
This duty-cycle estimate, combined with the previously derived energetics of the most recent outburst, yield an upper bound to the time-averaged rate of energy injection by the AGN of $\sim$$5\times10^{36}\rm\,W$.

\section{CONCLUSIONS}
\label{sec:CONCLUSIONS}

I have presented an extension to the RAiSE radio source evolution model \citep{Turner+2015, Turner+2017a} to quantify the energetics and duty-cycles of both remnant and restarted sources. The spectral indices of these objects are simulated over two typically observed frequency ranges ($151\rm\, MHz$ to $1.4\rm\, GHz$ and $5.5$ to $9.5\rm\, GHz$) to aid in their identification in large-sky surveys. Restarted and intermittent radio sources are found to have an integrated spectrum with a `steep-shallow' shape when the previous outburst occurred within $100\rm\, Myr$. By contrast, radio sources with a typical `shallow-steep' synchrotron spectrum generally have not had an outburst for over $1\rm\, Gyr$, or the lobes from any previous outbursts have completely mixed with the environment. The RAiSE synchrotron emissivity model also predicts that the weakest remnant sources will remain bright long after the jet switches-off; most of the energy in these objects is stored in the oldest synchrotron-emitting electrons. 

The combination of spectral ageing and radio source dynamical models is used to derive the intrinsic properties of the well-studied remnant radio galaxy B2\,0924+30. This remnant is found to have an age of $78_{-12}^{+14}\rm\, Myr$, have spent $50_{-9}^{+12}\rm\, Myr$ in the active phase with a jet kinetic power of $3.6_{-1.7}^{+3.0}\times10^{37}\rm\,W$, and have a lobe magnetic field strength below equipartition at the $8\sigma$ level. These measurements of the source age and active timescale are consistent with those of \citet{Jamrozy+2004} and \citet{Shulevski+2017}. Moreover, the energetics of B2\,0924+30 cannot be accurately reproduced without observations of the radio spectral shape. The time between outbursts with a comparable energy output to the observed remnant is found to be at least $330\rm\, Myr$, corresponding to a duty-cycle of $\delta < 0.15$ ($2\sigma$ level). The time-averaged energy imparted by the AGN into its host galaxy and cluster environment over the course of several outbursts is thus determined to be less than $\sim$$5\times10^{36}\rm\,W$.

The RAiSE radio source evolution model will be applied to {large samples} of remnants and restarted sources to determine typical AGN jet powers, active times and duty-cycles, enabling the energetics of AGN feedback to be quantified across the galaxy population for the first time.

\subparagraph{}
RJT thanks the CSIRO for a CASS studentship and Dr Stanislav Shabala for useful discussions. {I thank an anonymous referee for their insightful comments that have improved the manuscript.}

\end{document}